\documentclass[a4paper,12pt]{article}
\usepackage[utf8]{inputenc}
\usepackage[english]{babel}
\usepackage{amsmath, amssymb,graphicx,bm}
\usepackage{braket}
\usepackage{caption}
\usepackage{subcaption}
\usepackage{ulem}
\usepackage{multicol}
\usepackage{appendix}
\pdfoutput=1
\usepackage{jheppubm}
\newcommand{\be}{\begin{equation}}
\newcommand{\ee}{\end{equation}}
\newcommand{\bea}{\begin{eqnarray}}
\newcommand{\eea}{\end{eqnarray}}

\newcommand{\cS}{\mathcal{S}}
\newcommand{\cA}{\mathcal{A}}

\newcommand{\fb}{\mathfrak{b}}
\newcommand{\fc}{\mathfrak{c}}

\newcommand{\fg}{\mathfrak{g}}

\newcommand{\fz}{\mathfrak{z}}

\newcommand{\fG}{\mathfrak{G}}

\definecolor{darkraspberry}{rgb}{0.53,0.15,0.34}

\definecolor{darkblue}{rgb}{0,0,1}

\definecolor{dgreen}{rgb}{0,0.6,0}

\definecolor{brown}{rgb}{0.59,0.29,0}

\definecolor{orange}{rgb}{0.89,0.42,0.05}

\numberwithin{equation}{section}

\begin{document}

\title{ Holographic QCD Running Coupling  for Light Quarks in Strong Magnetic Field}

\author{Irina Ya. Aref'eva$^a$, Ali Hajilou$^a$, Alexander Nikolaev$^{a}$, Pavel Slepov$^a$}

\affiliation{$^a$Steklov Mathematical Institute, Russian Academy of  Sciences, \\ Gubkina str. 8, 119991, Moscow, Russia}

\emailAdd{arefeva@mi-ras.ru}
\emailAdd{hajilou@mi-ras.ru}
\emailAdd{nikolaev@mi-ras.ru}
\emailAdd{slepov@mi-ras.ru}

\abstract{We study the influence of magnetic field on the running coupling constant using  a bottom-up holographic model. We use the boundary condition that ensures the agreement with lattice calculations of string tension between quarks at zero chemical potential. The location of the 1st order phase  transitions in $(\mu, T)$-plane does not depend on the dilaton boundary conditions. We observe that the running coupling $\alpha$  decreases with increasing magnetic field for the fixed values of chemical potential and temperature.  At  the 1st order phase  transitions   the functions $\alpha$  undergo jumps depending on temperature, chemical potential and magnetic field. 
}
  
%\begin{verbatim}
   % Overleaf name Light quarks manuscript
%\end{verbatim}
%\begin{verbatim}
    %File name Aniz_AdS5_Light-Quarks_Magnetic_Field+plots.tex
%\end{verbatim}


\maketitle

\newpage

%\tableofcontents

\section{Introduction}
The goal of this paper is to study the running coupling in the presence of a strong external magnetic field. This work generalizes our previous investigation of the running coupling in isotropic holographic QCD \cite{Arefeva:2024vom} to the anisotropic QCD case, induced by an external magnetic field \cite{Arefeva:2020vae,Arefeva:2022avn}. We focus here specifically on the light-quark model \cite{Arefeva:2022avn}.
\\

The running coupling in QCD can be experimentally determined over a wide range of energies. These experimental data primarily pertain to scenarios with low density (small baryon chemical potential), such as those explored at the Large Hadron Collider (LHC) and the Relativistic Heavy Ion Collider (RHIC). The most recent experimental data can be found in \cite{ParticleDataGroup:2022pth,Deur:2022msf,Deur:2016tte,Brodsky:2010ur}. Our previous holographic results \cite{Arefeva:2024vom} can be compared with these experimental findings. For related studies using different holographic models, see \cite{IArefeva,Pirner:2009gr,Galow:2009kw,Brodsky:2010ur,Deur:2023dzc}. We use the holographic approach that can cover different ranges of energy from IR to UV domains, as an example see  \cite{Galow:2009kw}. Although, to cover all known experiments for IR and UV domains \cite{ParticleDataGroup:2022pth,Deur:2022msf} improved holographic models are needed.
In this study, we investigate the dependence of the running coupling on the holographic coordinate \( z \), which is related to the energy scale $E$ in boundary field theory (QCD) via the warp factor\footnote{
The energy scale $E$  in QFT can vary across a wide range and differs from the energy scale of QCD, denoted as 
$\Lambda_{QCD}$, which is fixed at the confinement scale, specifically 
$\Lambda_{QCD}=264$ MeV \cite{Galow:2009kw}.}. \\

In particular, it would be of interest to consider holographic isotropic models of both light and heavy quarks to study the running coupling as a function of the energy scale $E$, and to attempt to fit the results of these holographic models with experimental data such as 
those in \cite{Galow:2009kw,Pirner:2009gr}. The dilaton field plays a crucial role in the holographic approach to studying the running coupling, and it would be valuable to explore the effects of different boundary conditions on the running coupling as a function of the energy scale $E$ for both light and heavy quarks. Additionally, the holographic approach, combined with a special form of Fourier transformation from the 5-dimensional AdS space-time on the gravity side (\( z \) being the holographic coordinate) to the 4-dimensional Minkowski one, was used in \cite{Brodsky:2010ur} to study the behavior of \( \alpha \) as a function of momentum transfer \( Q^2 \). We plan to apply this approach to our model in future studies. In this paper, to obtain the dependence of the running coupling on the energy \( E \), we use a simpler relation provided by the prefactor in the 5-dimensional metric \cite{Pirner:2009gr,Galow:2009kw,IArefeva}.
\\

Quark-gluon plasma (QGP), a new phase of matter, is produced and studied in heavy ion collisions (HIC) at RHIC, LHC, and Nuclotron-based Ion Collider fAcility (NICA) experiments (see, for example, \cite{Du:2024wjm}). QGP is strongly coupled, and traditional methods, such as perturbation theory, are inadequate for studying its properties. Furthermore, lattice calculations encounter the sign problem at non-zero chemical potentials. Gauge/gravity duality offers a powerful non-perturbative approach for studying the strongly coupled regime of QCD \cite{Maldacena:1997re,Casalderrey-Solana:2011dxg, Arefeva:2014kyw,deWolf}. The structure of the QCD phase diagram is a critical area of research and can be explored within the holographic framework.
\\

Several experimental methods exist for studying the QCD phase transition diagram. One such method is the Beam Energy Scan (BES) \cite{STAR:2013cow,STAR:2005gfr}, which involves scanning a range of collision energies to search for signs of a 1st order phase transition, such as changes in particle yields or the onset of collective phenomena. Nonmonotonic behavior in both beam energy and impact parameter dependencies, if observed, can indicate a phase transition.
\\

Another method to identify a 1st order  phase transition is through the study of strange particle production. Anomalies in the production of strange particles, such as kaons and hyperons, may signal the presence of a 1st order phase transition \cite{STAR:2009sxc}. Analyzing higher-order correlations and cumulants of particle distributions can also provide evidence for critical phenomena associated with 1st order phase transitions \cite{Luo:2017faz}. Lastly, measuring elliptic flow as a function of collision energy can reveal changes in the medium's properties, potentially indicating a 1st order phase transition \cite{Huovinen:2006jp}.
\\

Here, we explore another possibility. As demonstrated in \cite{Arefeva:2024vom}, Holographic QCD (HQCD) predicts discontinuities in the running coupling's dependence on temperature \( T \) and chemical potential \( \mu \), resembling a 1st order phase transition in the \( (\mu, T) \)-plane. By identifying the locations of these discontinuities, we can predict the positions of 1st order phase transitions in the \( (\mu, T) \)-plane. This requires access to high-density matter, particularly at high chemical potentials, which might be achievable in future experiments such as NICA and  the Facility for Antiproton and Ion Research (FAIR).\\

During HIC, strong magnetic fields are generated \cite{Zhang:2023ppo,Skokov:2009qp,Bzdak:2011yy,Voronyuk:2011jd,Deng:2012pc}, which in turn affect the running coupling constant. Therefore, it is essential to assess the impact of strong magnetic fields on the running coupling in HQCD. In other words, we aim to determine how the results from \cite{Arefeva:2024vom} are modified in the presence of a magnetic field. This is the primary objective of this paper. As noted in \cite{Arefeva:2024vom}, the QCD phase diagram is significantly influenced by quark masses. For simplicity, we focus on the scenario involving light quarks.
\\

The typical behavior of the running coupling \( \alpha \) $(E; \mu, T)$, particularly the logarithm of the running coupling \( \log \alpha \) $(E; \mu, T)$, as predicted by holographic models, was presented in our previous work \cite{Arefeva:2024vom}. This behavior is shown at fixed energy scales, various temperatures, and chemical potentials (see also Fig.\,\ref{Fig:phi-cb0} below). A notable feature in different parts of this graph is the appearance of jumps in the coupling constant, indicating 1st order phase transitions along the same lines in the \( (\mu,T) \)-planes, irrespective of the energy scales. These jumps cease at the critical endpoint (CEP), marking the end of the critical points' lines. For non-zero magnetic fields, the 1st order phase transition persists up to a certain value of the magnetic field parameter \( c_B \) that $c_B\neq0$ is responsible for anisotropy in our  holographic model, as illustrated in Fig.\,\ref{Fig:Tmu} and Fig.\,\ref{Fig:cb005}.
\\

The paper is organized as follows. 
In Sect.\,\ref{sec:prelim}, we present the 5-dimensional holographic models in the presence of a non-zero magnetic field for light quarks.
In Sect.\,\ref{sec:running}, we describe the influence of the magnetic field on the running coupling constant in the light-quark model.
In Sect.\,\ref{Sect:Conc}, we summarize the results obtained.
The paper is supplemented by Appendix\,\ref{app1}, which details how the equations of motion (EOMs) were solved.

\section{Holographic Model} \label{sec:prelim}

Studying the effect of external magnetic field on the QCD features via holography is discussed in  \cite{Arefeva:2020vae,Arefeva:2022avn, Gursoy:2017wzz,Arefeva:2023ter,Bohra:2019ebj,Arefeva:2023jjh,Chen:2024jet}. To examine the behavior of the running coupling constant, we utilize a holographic model for light quarks in a magnetic field, which includes a 1st order phase transition \cite{Arefeva:2022avn,Arefeva:2022bhx}. The holographic heavy-quark models including two types of anisotropy have been proposed in \cite{Arefeva:2020vae,Arefeva:2023jjh}. These models extend the corresponding isotropic models without magnetic fields \cite{Yang:2015aia,Li:2017tdz}, and the anisotropic versions, which feature spatial anisotropy, are related to non-central HIC  \cite{Arefeva:2020byn,Arefeva:2014vjl,Arefeva:2018hyo,Arefeva:2018cli}.

\subsection{Background Setup} \label{sec:model}

We consider the 5-dimensional action in the Einstein frame with two Maxwell fields, given by:
\begin{gather}
  {\cS} = \int \cfrac{d^5x \, }{16\pi G_5} \sqrt{-\fg} \left[
    R - \cfrac{f_1(\varphi)}{4} \ F^{(1)^2}
        - \cfrac{f_B(\varphi)}{4} \ F^{(B)^2}
    - \cfrac{\partial_{\mu} \varphi \partial^{\mu} \varphi}{2}
    - V(\varphi) \right]\,, \label{eq:2.01}
\end{gather}
where $\fg$ is the determinant of the metric tensor, $G_5$ is the 5-dimensional Newtonian gravitational constant, and $R$ is the Ricci scalar. Here, we use the spacetime coordinates $(t, x, y_1, y_2, z)$, with $z$ being the holographic variable. The ansatz for the non-zero components of the first and second Maxwell fields are:
\begin{gather}
  A_{\mu}^{(1)} = A_t (z) \delta_\mu^0, \quad
  F_{x y_1}^{(B)} = q_B. \label{eq:2.02}
\end{gather}
In the action \eqref{eq:2.01}, $\varphi = \varphi(z)$ represents the dilaton field. The functions $f_1(\varphi)$ and $f_B(\varphi)$ are the coupling functions associated with the Maxwell fields $A_{\mu}$ and $F_{\mu\nu}^{(B)}$, respectively. Here, $q_B$ is a constant charge, and $V(\varphi)$ denotes the dilaton field potential. According to the holographic dictionary, $F^{(1)}$ and $F^{(B)}$ in the gravity background correspond to the chemical potential and magnetic field in the 
boundary field theory, respectively.
%\\

The actions used in \cite{Arefeva:2018hyo, Arefeva:2020byn} also include two Maxwell fields, where the second Maxwell field supports a spatial anisotropy. In contrast, in the action \eqref{eq:2.01} the second Maxwell field  supports an external magnetic field, $F_{\mu\nu}^{(B)}$, contributing to a different anisotropy.

Our ansatz for the metric is:
\begin{gather}
  ds^2 = \cfrac{L^2 \, \fb(z)}{z^2} \left[
    - \, g(z) dt^2 + dx^2 
    +  \, dy_1^2
    + e^{c_B z^2}  dy_2^2
    + \cfrac{dz^2}{g(z)} \right], \label{eq:2.03} \\
  \fb(z) = e^{2{\cA}(z)}\,,  \label{eq:2.04}
\end{gather}
where $L$ is the AdS radius and $\fb(z)$ is the warp factor. Here, we set $L=1$. To clarify the parameters of the metric in \eqref{eq:2.03}, note that the difference between the "heavy-quark" and "light-quark" cases lies in the form of the scale factor ${\cA}(z)$, in the warp factor. For the heavy-quark model a simple choice is ${\cA}(z)=-\,\fc\,z^2/4$ \cite{Arefeva:2018hyo}, where the parameter $\fc = 1.547$ GeV${}^2$ is fitted to experimental data \cite{Yang:2015aia} and for an extended scale factor one can see \cite{Arefeva:2023jjh}. For the "light-quark" case, following \cite{Li:2017tdz}, we assume:
\be\label{sfactor}
{\cA}(z) = - \, a \ln (b z^2 + 1)
\ee
where the parameters $a = 4.046$ and $b = 0.01613$ GeV${}^2$ can be  obtained by fit to experimental data \cite{Li:2017tdz}.

In the metric ansatz \eqref{eq:2.03}, $g(z)$ is the blackening function, which plays a crucial role in calculating the temperature and entropy of the black hole in our holographic model. Additionally, to describe the non-centrality of HIC, we introduce the magnetic field parameter, i.e. $c_B$. The relationship between $c_B$ and the magnitude of the magnetic field $B$ is given by $B^2 = - \, c_B$  \cite{Bohra:2019ebj, Bohra:2020qom, DHoker:2009ixq}. Therefore, in our calculations, we consider $c_B \leq 0$. For our numerical calculations, we generally set the charge $q_B$ to $q_B=1$ and vary the magnetic field parameter $c_B$. In the metric \eqref{eq:2.03}, the external magnetic field breaks the $SO(3)$ invariance in the boundary coordinates $(x, y_1, y_2)$. When the magnetic field is turned off, the $SO(3)$ invariance is restored.

By varying the action, the EOMs are obtained:

\begin{gather}
 \begin{split}
    \varphi'' &+ \varphi' \left( \cfrac{g'}{g} + \cfrac{3 \fb'}{2 \fb} -
      \cfrac{3}{z} + c_B z \right)
    + \left( \cfrac{z}{L} \right)^2 \cfrac{\partial f_1}{\partial
      \varphi} \ \cfrac{(A_t')^2}{2 \fb g} - \left( \cfrac{z}{L} \right)^{2} \cfrac{\partial
      f_B}{\partial \varphi} \ \cfrac{q_B^2}{2 \fb g} \\ &
    \qquad \qquad \qquad \qquad \qquad - \left( \cfrac{L}{z} \right)^2 \cfrac{\fb}{g} \ \cfrac{\partial
      V}{\partial \varphi} = 0\,,
  \end{split}
  \label{eq:2.05} \\
      A_t'' + A_t' \left( \cfrac{\fb'}{2 \fb} + \cfrac{f_1'}{f_1} -
    \cfrac{1}{ z} + c_B z \right) = 0\,, \label{eq:2.06} \\
    g'' + g' \left(\cfrac{3 \fb'}{2 \fb} - \cfrac{3}{ z} + c_B
    z \right)
  - \left( \cfrac{z}{L} \right)^2 \cfrac{f_1 (A_t')^2}{\fb}
  - \left( \cfrac{z}{L} \right)^{2} \cfrac{q_B^2 \
    f_B}{\fb} = 0\,, \label{eq:2.07}
\\  
  \fb'' - \cfrac{3 (\fb')^2}{2 \fb} + \cfrac{2 \fb'}{z}
 + \cfrac{2 \fb\, c_B}{3 } \left( 1
    + c_B z^2 \right)
  + \cfrac{\fb \, (\varphi')^2}{3} = 0\,, \label{eq:2.08} 
  \\
  \begin{split}
   c_B z^2  \left(2 g'+3 g \right) 
      \left(
        \cfrac{\fb'}{\fb} - \cfrac{4}{3 z} + \cfrac{2 c_B z}{3}
      \right) 
    - \left( \cfrac{z}{L} \right)^{3}
    \cfrac{L \,q_B^2 f_B }{\fb} = 0\,, 
  \end{split}\label{eq:new} \\
  \begin{split}
    \cfrac{\fb''}{\fb} &+ \cfrac{(\fb')^2}{2 \fb^2}
    + \cfrac{3 \fb'}{\fb} \left( \cfrac{g'}{2 g}
      - \cfrac{2}{ z}
      + \cfrac{2 c_B z}{3} \right)
    - \cfrac{g'}{3 z g}  \left(9 - 3 c_B z^2
    \right)  - \cfrac{2 c_B}{3} \left(  5-c_B z^2 \right)\\
    & \qquad \qquad \qquad \qquad \quad + \cfrac{8}{ z^2} 
    + \cfrac{g''}{3 g} + \cfrac{2}{3} \left( \cfrac{L}{z} \right)^2
    \cfrac{\fb V}{g} = 0\,,
  \end{split}\label{eq:2.10}
\end{gather}
%%%%%%%%%%%%%%%%%%%%%%%%%%%%%
%\begin{verbatim}
%   Action_vary_full_qB12.nb
%\end{verbatim}
%%%%%%%%%%%%%%%%%%%%%%%%%%%%%
Here, the symbol "$'$" denotes differentiation with respect to the holographic coordinate $z$. The EOMs \eqref{eq:2.05}–\eqref{eq:2.10} have a general form and can be applied to both heavy and light-quark models.

To solve the EOMs, we apply the following boundary conditions:
\begin{gather}
  A_t(0) = \mu, \quad A_t(z_h) = 0, \label{eq:2.11} \\
  g(0) = 1, \quad g(z_h) = 0, \label{eq:2.12} \\
  \varphi(z_0) = 0\, , \label{eq:2.13}
\end{gather}
where $z_h$ denotes the size of the black hole horizon and $z_0$ is an arbitrary value of the holographic coordinate $z$. Different choices for the boundary condition of the dilaton field $\varphi$ have been discussed in detail in \cite{Arefeva:2024vom, Li:2017tdz, Arefeva:2020byn, Arefeva:2018hyo}. For instance, \cite{Li:2017tdz} uses $z_0 = 0$, while \cite{Arefeva:2018hyo} adopts $z_0 = z_h$.

%%%%%%%%%%%%%%%%%%%%%%%%%%
%\begin{verbatim}
%    Action_vary_HL.nb
%\end{verbatim}
%%%%%%%%%%%%%%%%%%%%%%%%%%

\subsection{Thermodynamics and Background Phase Transition} \label{Sect:therm}

\begin{figure}[h!]
  \centering
\includegraphics[scale=0.16]{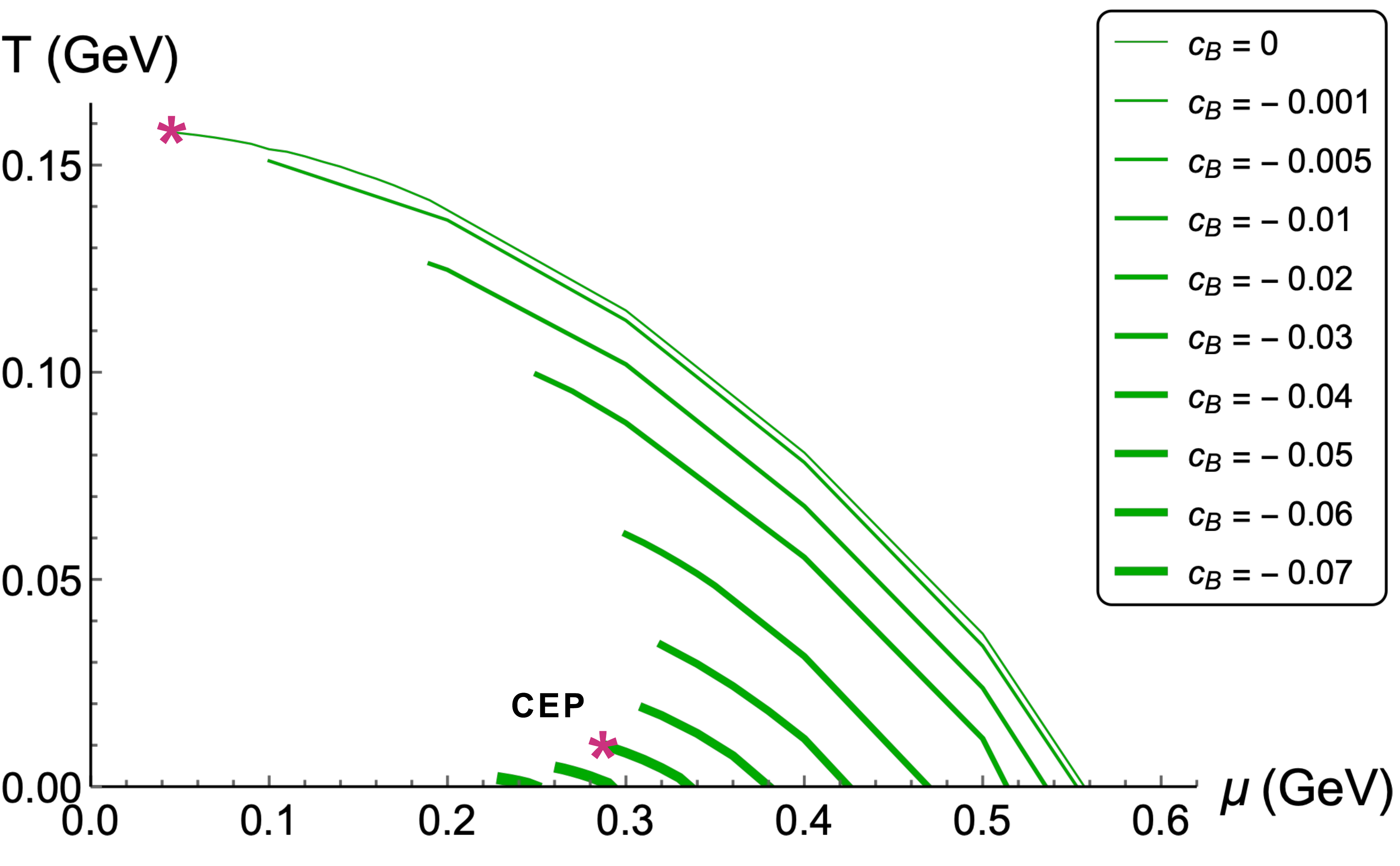} 
\caption{Phase diagram in the \((\mu,T)\)-plane for the light-quark model, showing the isotropic case \(c_B = 0\) and anisotropic cases with different \( c_B \). The magenta stars indicate the CEPs; \( [c_B] = \) GeV${}^2$.}
  \label{Fig:Tmu}
\end{figure}

For the metric in   \eqref{eq:2.03}, the temperature and entropy for the light-quark model can be expressed as:
\begin{gather}
  T = \cfrac{|g'|}{4 \pi} \, \Bigl|_{z=z_h}\,, \qquad
  s = \left( \cfrac{L}{z_h} \right)^{3}
  \cfrac{e^{c_B z_h^2/2}}{4 \left( 1 + b z_h^2
    \right)^{3a}} \,, \label{eq:3.03}
\end{gather}
where in the entropy formula, we set \( G_5 = 1 \). The phase diagrams for the light-quark model can be obtained through free energy considerations:
\begin{gather}
  F = - \int s \, d T = \int_{z_h}^{\infty} s \, T' \, dz, \label{eq:3.05}
\end{gather}
where the free energy is normalized to vanish as \( z_h \to \infty \). The phase transition curves for the background become shorter as the magnetic field increases (i.e., for larger absolute values of \( c_B \)). Using the free energy from  \eqref{eq:3.05}, the phase diagram in the \((\mu,T)\)-plane for the light-quark model is shown in Fig.\,\ref{Fig:Tmu}, illustrating both the isotropic case (\(c_B = 0\)) and anisotropic cases with different \( c_B \). The corresponding critical end points (CEPs) are marked with magenta stars.

\begin{figure}[t!]
  \centering
\includegraphics[scale=0.46]{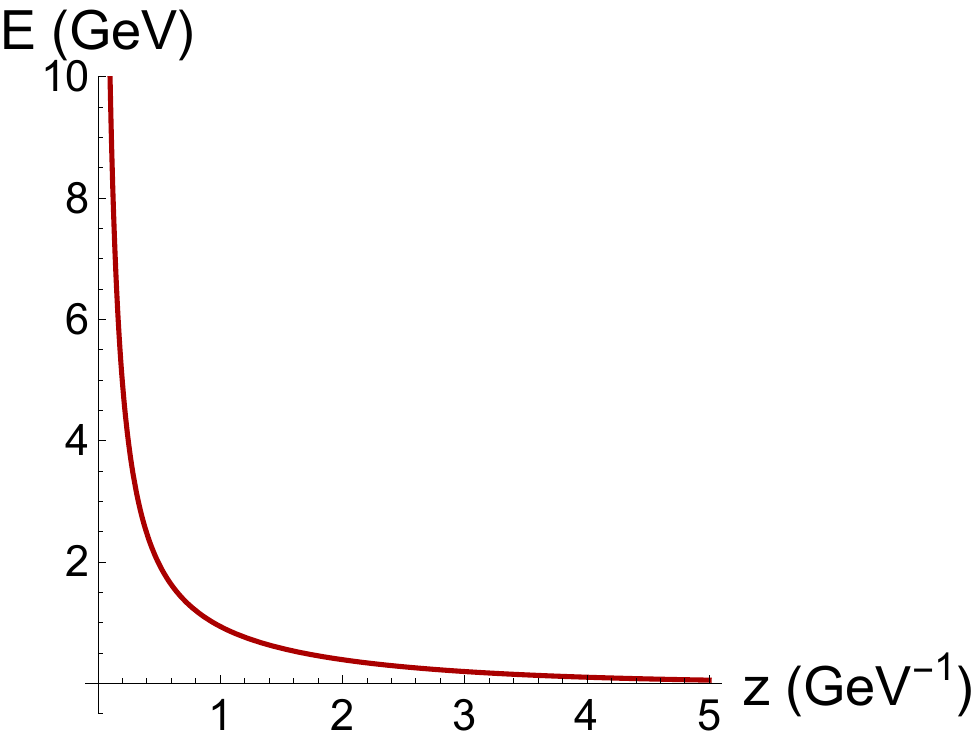} 
\caption{Energy scale \( E \) (GeV) in the boundary field theory as a function of the holographic coordinate \( z \) (GeV${}^{-1}$), corresponding to the warp factor \( \fb(z) \).}
 \label{Fig:E-z-LQ}
\end{figure}

The energy scale \( E \) (GeV) in the boundary field theory as a function of the holographic coordinate \( z \) (GeV${}^{-1}$), corresponding to the prefactor of the metric \eqref{eq:2.03}, is defined as \cite{Galow:2009kw}:
\be \label{BBL}
E = \frac{\sqrt{\fb(z)}}{z} = \frac{\left(b z^2 + 1\right)^{-a}}{z}\, ,
\ee
where the parameters \( a \) and \( b \) were introduced in \eqref{sfactor}.
The energy scale \( E(z) \) is depicted in Fig.\,\ref{Fig:E-z-LQ}.

\begin{figure}[h!]
  \centering
     \includegraphics[scale=0.7]{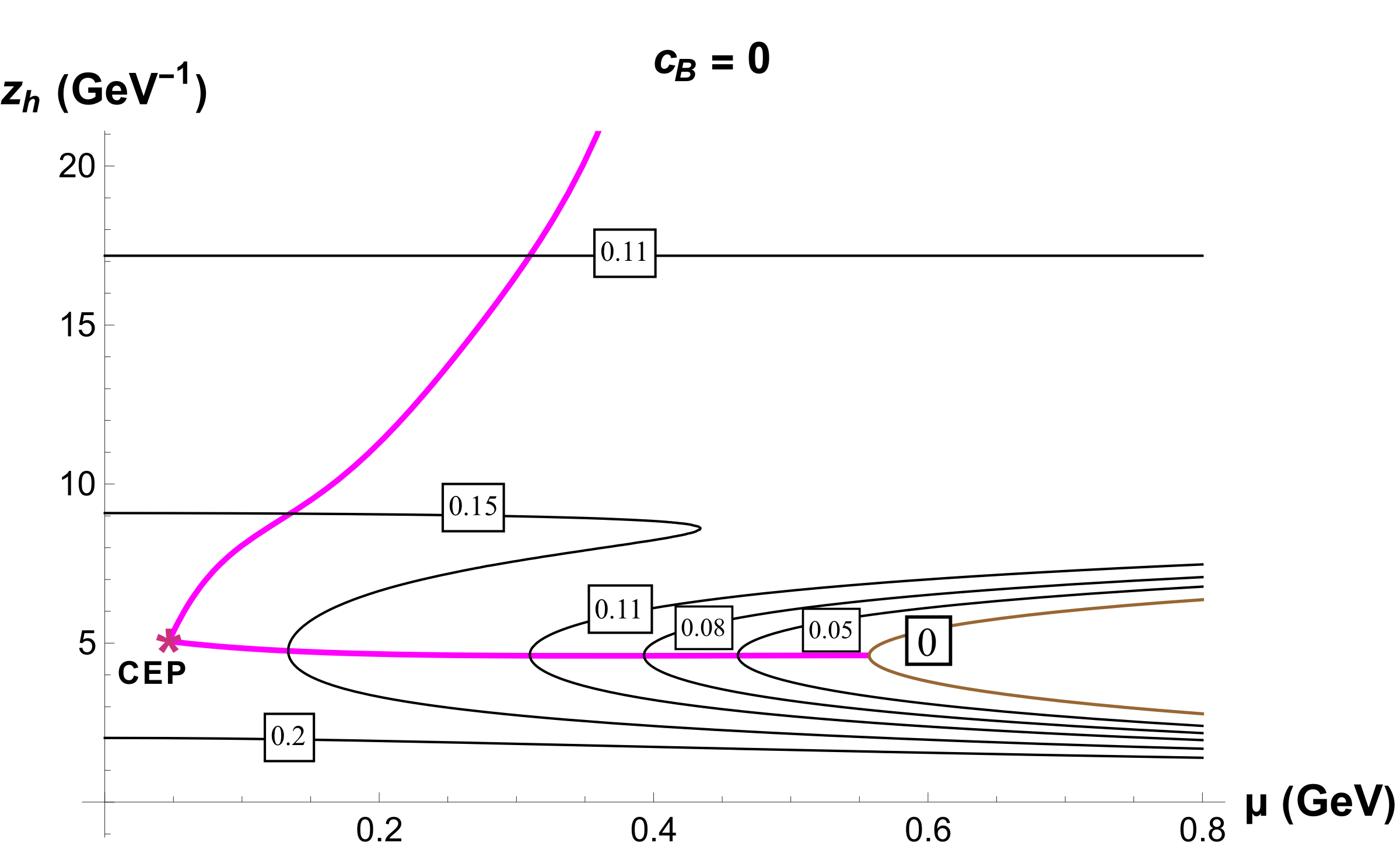} 
    \\{\bf A}\\\,\\
  \includegraphics[scale=0.7]{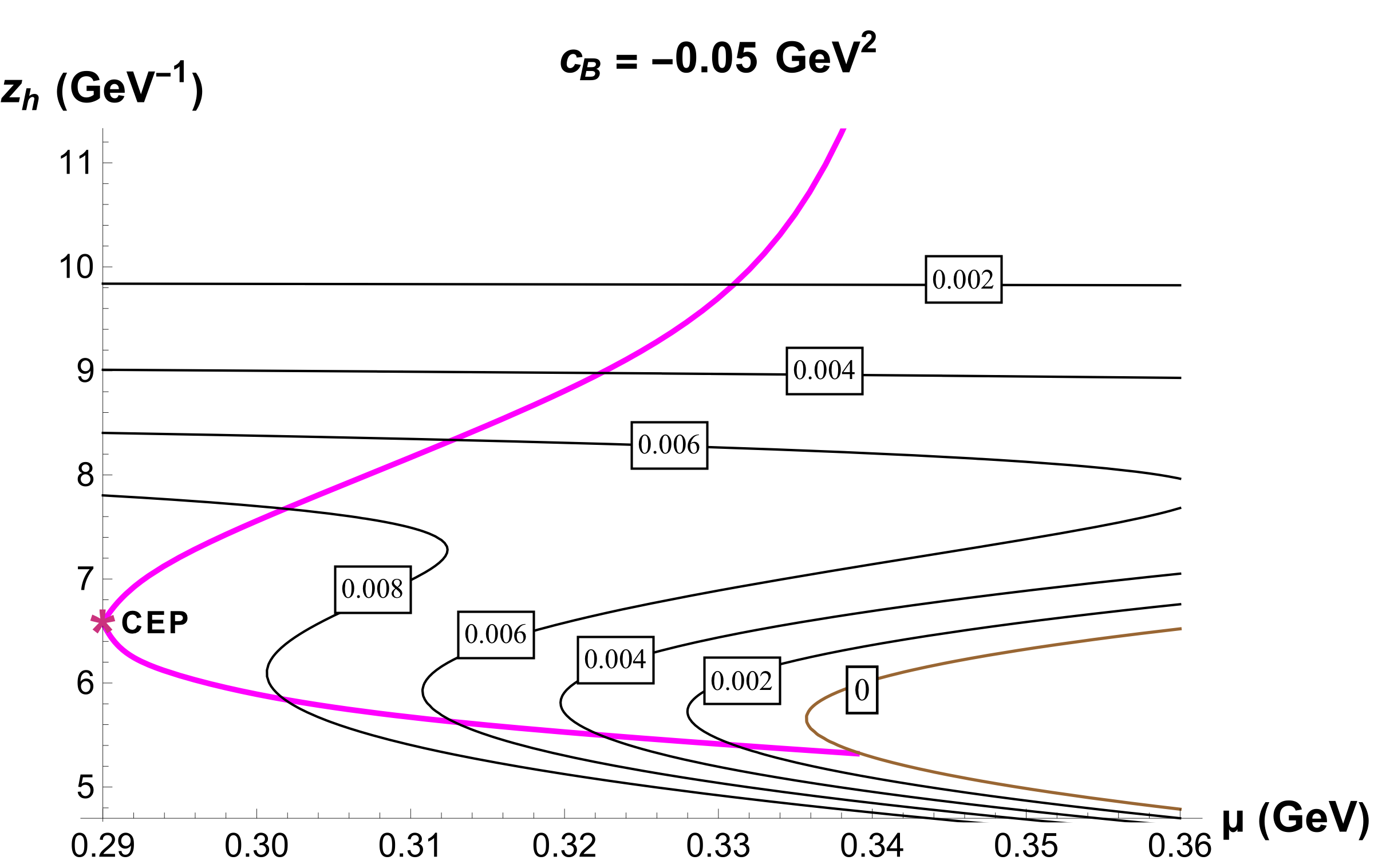} \\ {\bf B} 
  \caption{2D plots in the \((\mu,z_h)\)-plane for \( c_B = 0 \) (A) and \( c_B = -0.05 \) GeV${}^2$ (B). The magenta curves show the coordinates of the 1st order phase transition; the black contours correspond to fixed temperatures, and the brown contour represents \( T = 0 \). The magenta stars indicate the CEPs; \( [T] = \) GeV.}
  \label{Fig:Z_h mu approx}
\end{figure}

Fig.\,\ref{Fig:Z_h mu approx} shows 2D plots in the \((\mu,z_h)\)-planes for \( c_B = 0 \) (A) and \( c_B = -0.05 \) GeV${}^2$ (B). The magenta curves indicate the coordinates of the 1st order phase transition, the black contours correspond to fixed temperatures, and the brown contour corresponds to the \( T = 0 \) case. The magenta star represents the CEP. Each magenta curve consists of two branches—upper and lower—connected at the CEP. The regions of stable black hole solutions are located above the upper magenta curve and below the lower branch, corresponding to physical regions, while the area between the branches corresponds to an unstable or nonphysical region. The region above the upper branch describes the hadronic phase, and the region below the lower branch corresponds to the QGP phase.

\section{Running Coupling in a Strong Magnetic Field}\label{sec:running}

The holographic running coupling \( \alpha(z) \) is defined in terms of the dilaton field \( \varphi(z) \) as \cite{Gursoy:2007cb,Gursoy:2007er,Pirner:2009gr}:
\be\label{lambda-phi}
\alpha(z) = e^{\varphi(z)} \,.
\ee
It is important to note that defining the dilaton field \( \varphi(z) \) requires specifying a boundary condition. One option is the zero boundary condition at \( z = 0 \):
\be \label{zerobc1}
\varphi_0(z)\Big|_{z=0} = 0\,.
\ee
Alternatively, a boundary condition can be applied at an arbitrary holographic coordinate \( z_0 \):
\be \label{zerobc2}
\varphi_{z_0}(z)\Big|_{z=z_0} = 0\,.
\ee
The relationship between these two solutions is given by:
\be
\varphi_{z_0}(z) = \varphi_0(z) - \varphi_0(z_0)\,.
\ee
Using this relation, the running coupling can be expressed as \cite{Arefeva:2024vom}:
\be
\alpha_{z_0}(z) = \alpha_0(z) \, \fG(z_0)\,, \quad \alpha_0(z) = e^{\varphi_0(z)}\,, \quad \fG(z_0) = e^{-\varphi_0(z_0)}\,.
\ee

It is important to note that these boundary conditions, \eqref{zerobc1} and \eqref{zerobc2}, do not depend on \( z_h \), and therefore, they cannot be used to study the thermodynamics of the system. To explore the dependence of the running coupling on thermodynamic quantities such as \( T \) and \( \mu \), we propose a new boundary condition, \( z_0 = \fz(z_h) \). With this condition, the running coupling is given by:
\be
\alpha_{\fz}(z; T, \mu) = \alpha_0(z) \, \fG(T, \mu)\,, \qquad \fG(T, \mu) = e^{-\varphi_0(\fz(z_h))}\,.
\ee
For the new boundary condition, one option is to choose \( \fz(z_h) = z_h \). Another choice, given by an exponential function, which defines a new boundary condition for the light-quark model \cite{Arefeva:2024vom}, is:
\begin{equation}\label{LQz0}
z_0 = \fz(z_h) = 10 \, \exp\left(-\frac{z_h}{4}\right) + 0.1 \,, 
\end{equation}
where the dilaton field is zero at \( z_0 \), as indicated in \eqref{eq:2.13}. It is important to note that the boundary condition in \eqref{LQz0}, that was introduced in \cite{Arefeva:2020byn}, was derived by studying the behavior of the QCD string tension as a function of the temperature at zero chemical potential for the isotropic case determined by lattice calculations \cite{Cardoso:2011hh}. Although lattice results are not available for the anisotropic case (\( c_B \neq 0 \)), this boundary condition is still used in the present research.

To study the behavior of the running coupling constant for magnetic fields \( c_B = 0 \) and \( c_B = -0.05 \) GeV${}^2$, we refer to Fig.\,\ref{Fig:Z_h mu approx} to identify the physical regions of the model, which are limited by the sizes of the black hole horizons, \( z_h \), corresponding to the 1st order phase transition (magenta line).

\begin{figure}
  \hspace*{-0.7cm}
  \includegraphics[scale=0.18]{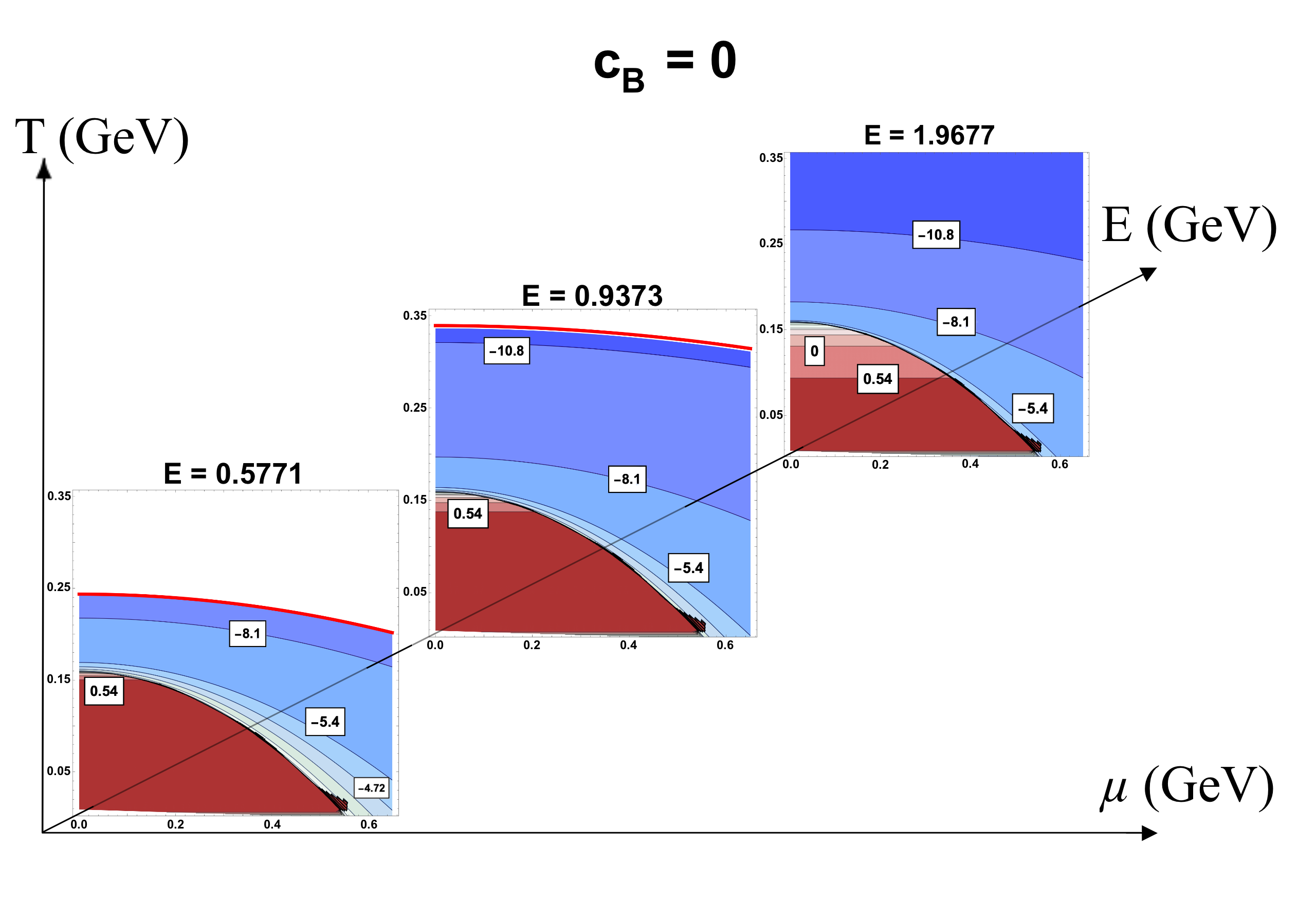}
  \caption{Density plots with contours for \( \log\alpha_{\fz}(E; \mu, T) \) at different energy scales \( E = \{0.5771, 0.9373, 1.9677 \} \) (GeV) for \( c_B = 0 \). All values of $E$ on the top of each panel show fixed value of energy  $E$-coordinate.  
  The red lines in the first two graphs on the left indicate the maximum possible temperature values at the corresponding energy scale \( E \).}
\label{Fig:phi-cb0}
\end{figure}

In Fig.\,\ref{Fig:phi-cb0}, density plots with contours for $\log\alpha_{\fz}(E; \mu, T)$  at different energy scales $ E = \{0.5771, 0.9373, 1.9677\}$  (GeV) for \( c_B = 0 \) are presented as a family of sections in three-dimensional space \( (E, \mu, T) \). The selected energy scales correspond to holographic coordinates \( z = \{0.5, 1, 1.5\} \) (GeV${}^{-1}$) as shown in Fig.\,\ref{Fig:E-z-LQ}. Each plot includes several contours corresponding to fixed values of \( \log\alpha_{\fz} \), with the values are indicated in the rectangles. The red lines in the first two graphs on the left denote the maximum possible temperature values at the corresponding energy scales \( E \). These constraints arise due to the limitation \( z \leq z_h \). The plot in Fig.\,\ref{Fig:phi-cb0} is a reproduction of a plot from Ref.\,\cite{Arefeva:2024vom}, i.e.  Fig.\,23. It is evident that as the energy scale increases, the contours for fixed values of the running coupling shift to lower temperatures. The running coupling decreases monotonically with increasing temperature, indicating that the running coupling constant decreases as the energy scale increases for a fixed given temperature.

\begin{figure}[t!]
  %\centering
  \hspace*{-0.7cm}
  \includegraphics[scale=0.2]{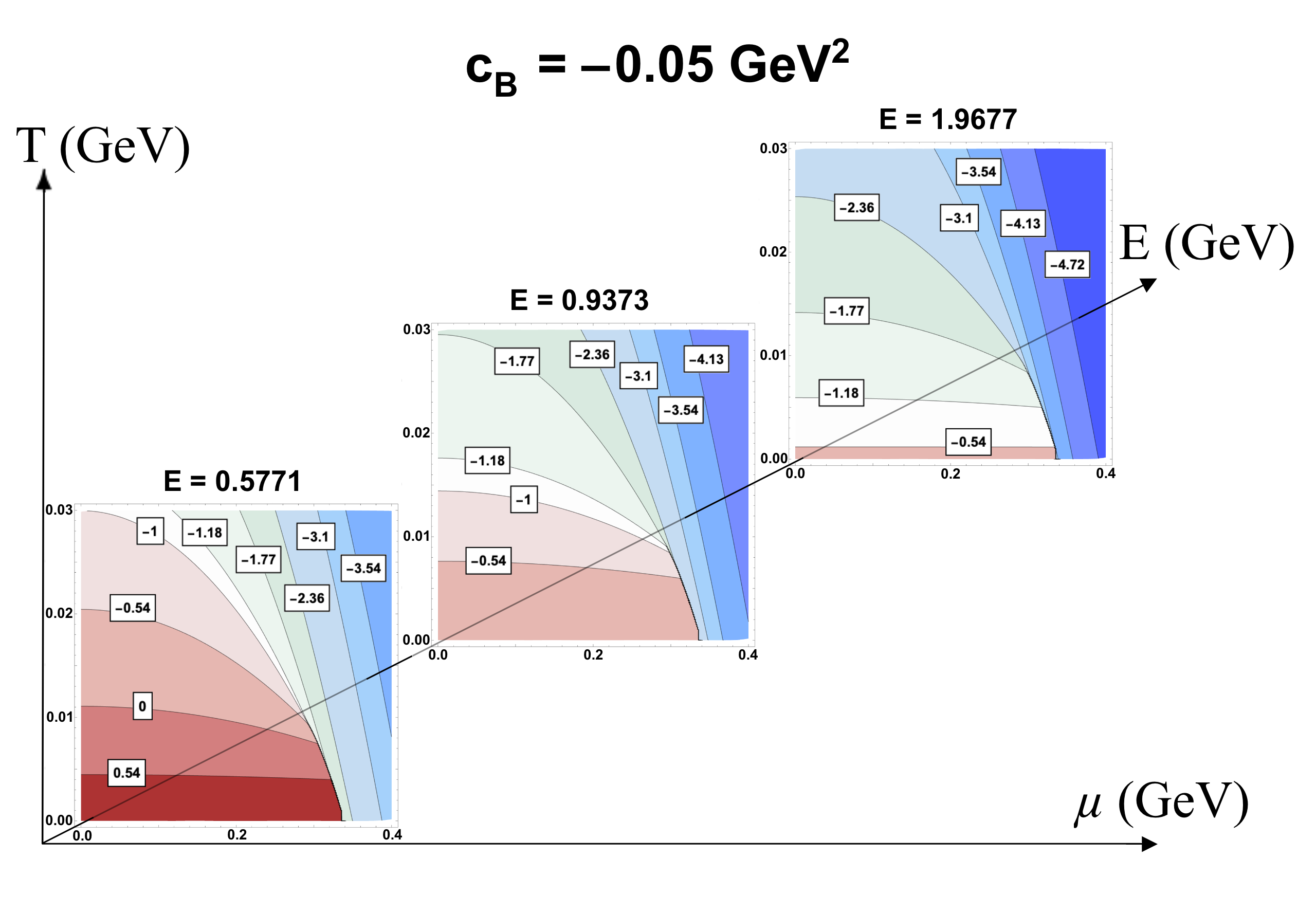}
  \caption{Density plots with contours for \( \log\alpha_{\fz}(E; \mu, T) \) at different energy scales \( E = \{0.5771, 0.9373, 1.9677\} \) (GeV) are shown for \( c_B = -0.05 \) GeV${}^2$. All values of $E$ on the top of each panel show fixed value of energy  $E$-coordinate. These plots illustrate the variation of the running coupling constant across different energy scales, chemical potentials, and temperatures in the presence of a magnetic field.}
\label{Fig:cb005}
\end{figure}

Fig.\,\ref{Fig:cb005} presents density plots with contours for \( \log\alpha_{\fz}(E; \mu, T) \) the same as Fig.\,\ref{Fig:phi-cb0}, but with the magnetic field turned on, \( c_B = -0.05 \) GeV${}^2$. 

In the isotropic case depicted in Fig.\,\ref{Fig:phi-cb0}, the running coupling constant in the hadronic phase shows minimal dependence on the chemical potential. In contrast, for the anisotropic case with \( c_B = -0.05 \) GeV${}^2$ shown in Fig.\,\ref{Fig:cb005}, there is a slight dependence of the running coupling on the chemical potential in the hadronic phase. However, in QGP phase for both cases, i.e. \( c_B = 0 \) and \( c_B = -0.05 \) GeV${}^2$, the running coupling exhibits significant dependence on the chemical potential and temperature.

Comparing the plots in Fig.\,\ref{Fig:phi-cb0} and Fig.\,\ref{Fig:cb005} for fixed parameters \((E, \mu, T)\), it is evident that the introduction of a magnetic field decreases the value of the running coupling constant.

\newpage

\section{Conclusion} \label{Sect:Conc}

In this paper, we have examined an anisotropic holographic model with a strong magnetic field for the light-quark case \cite{Arefeva:2022avn,Arefeva:2022bhx}. This model is characterized by the Einstein-dilaton-two-Maxwell action and a 5-dimensional metric with a warp factor, which was previously considered in the isotropic case for light quarks \cite{Li:2017tdz}. Our model specifically incorporates anisotropy due to the external magnetic field represented by the second Maxwell field.

It is crucial to note that different boundary conditions lead to varying physical outcomes. The boundary condition given by \eqref{LQz0} for light quarks aligns with lattice calculations for non-zero temperature and zero chemical potential \cite{Kaczmarek:2007pb}.

To summarize our findings:

\begin{itemize}
    \item The magnetic field reduces the value of the running coupling constant \(\alpha\) at fixed temperatures \(T\) and chemical potentials \(\mu\).
    \item The running coupling \(\alpha\) decreases with increasing energy scale \(E\) at fixed \(T\) and \(\mu\), consistent with the well-known asymptotic freedom, and this behavior persists even in the presence of a magnetic field.
    \item In the hadronic phase, the running coupling \(\alpha\) varies slowly with changes in chemical potential, both for the isotropic case (\(c_B = 0\)) and the anisotropic case (\(c_B \neq 0\)).
    \item 
     In the QGP phase, the running coupling \(\alpha\) exhibits significant dependence on the chemical potential and temperature,  both for the isotropic case (\(c_B = 0\)) and the anisotropic case (\(c_B \neq 0\)).
    \item Along the 1st order phase transition line, the running coupling exhibits a discontinuity. This discontinuity decreases along the transition line and vanishes at the critical end point (CEP).
\end{itemize}

We would like to emphasize that the twice anisotropic model for light quarks is significantly more complex than the twice anisotropic model for heavy quarks \cite{Arefeva:2020vae}, due to more complicated forms of the deformation factor and the kinetic gauge function \(f_{1}\). The effect of the magnetic field on the running coupling \(\alpha\) in the heavy-quark model will be explored in a future research.

\section{Acknowledgments}

We thank K. Runnu and M. Usova for their valuable discussions. The work of I.A., P.S., and A.N. is supported by the Russian Science Foundation grant No. 20-12–00200. The work of A.H. was conducted at the Steklov International Mathematical Center and supported by the Ministry of Science and Higher Education of the Russian Federation (Agreement No. 075-15-2022-265). We also appreciate the referee’s comments, which helped to improve the presentation of this manuscript.

\newpage
\appendix
{\large \bf Appendix}
%\begin{appendices}
%\section{Solution to EOMs}
%\end{appendices}

\section{Solution to EOMs }\label{app1}

The system of EOMs \eqref{eq:2.05}--\eqref{eq:2.10} contain six equations, although just five of them are independent. In other words, Equ. \eqref{eq:2.05} is the consequence of \eqref{eq:2.06}--\eqref{eq:2.10}. To solve EOMs
\eqref{eq:2.05}--\eqref{eq:2.10} we need to choose the form of the 
coupling function $f_1$ \cite{Arefeva:2022avn}. Taking into account the ``light-quark'' scale factor, we get
\begin{gather}
  f_1 = e^{-cz^2-{\cA}(z)} 
  = (1 + b z^2)^a \, e^{- c z^2} \, , \label{eq:2.14}
\end{gather}
 where $c=0.227$   GeV${}^2$, $a$ and $b$ have already been introduced in \eqref{sfactor}. Solving \eqref{eq:2.06} with the coupling function \eqref{eq:2.14} and
boundary conditions \eqref{eq:2.11} gives
\begin{gather}
  A_t = \mu \, \cfrac{e^{\left(2c-c_B\right)z^2/2} -
    e^{\left(2c-c_B\right)z_h^2/2}}{1 -
    e^{\left(2c-c_B\right)z_h^2/2}}\,. \label{eq:2.15} 
\end{gather}

We get an expression for the  function $f_B$
\begin{gather}
  f_B = 2 \left( \cfrac{z}{L} \right)^{-2} \!\!\! \fb g \,
  \cfrac{c_B z}{q_B^2}
  \left( \cfrac{3 \fb'}{2 \fb} - \cfrac{2}{z} + c_B z +
    \cfrac{g'}{g} \right), \label{eq:2.17}
\end{gather}
and get the solution for blackening function:
\begin{gather}
  g = e^{c_B z^2} \left[ 1 - \cfrac{I_1(z)}{I_1(z_h)}
    + \cfrac{\mu^2 \, (2 c - c_B) \, I_2(z)}{L^2 \left( 1 -
        e^{\left(2c-c_B\right)z_h^2/2} \right)^2} \left( 1 -
      \cfrac{I_1(z)}{I_1(z_h)} \, \cfrac{I_2(z_h)}{I_2(z)} \right)
  \right], \label{eq:2.20}
\end{gather}
where 
\begin{gather}
 I_1(z) = \int_0^{z} \left(1 + b \xi^2 \right)^{3a} e^{- 3 c_B
      \xi^2/2} \, \xi^{3} \, d \xi, \\
  I_2(z) = \int_0^{z} \left(1 + b \xi^2 \right)^{3a}
  e^{\left(c-2c_B\right)\xi^2} \, \xi^{3} \, d
  \xi. \label{eq:2.22}
\end{gather} 

Further solving \eqref{eq:2.05}--\eqref{eq:2.10} system with the coupling
functions \eqref{eq:2.14} and \eqref{eq:2.17} and the boundary condition
\eqref{eq:2.13} 
gives

\begin{gather}
   \varphi = \int_{z_0}^z \sqrt{ - 2 c_B - 2 c_B^2 \xi^2
    + \cfrac{12 a b}{1 + b \xi^2} \left( 1 + 2 \, \cfrac{1 + a b
        \xi^2}{1 + b \xi^2} \right) } \, d \xi, \label{eq:2.24} \\
  \begin{split}
    &V(z) = - \, \cfrac{(1 + b z^2)^{2a} \ z}{2 L^2} \ \times 
     \left[ 2 \left\{ 2 \, \cfrac{6 + \bigl( 2 + 5(2 + 3a ) \bigr) b z^2
     + (2 + 3 a) (3 + 6 a ) b^2 z^4}{(1 + b z^2)^2  z^2} \right.  \right. \\
      & \left. - \left. c_B \left(5+ \cfrac{12 a b z^2}{1 + b z^2} \right) + c_B^2 z^2
        \right\} g \, z
        - \left( 5 + 3 \left( \cfrac{6 a b}{1 + b z^2} - c_B \right)
          z^2 + 4 \right) g' + g'' z \right].
  \end{split}\label{eq:2.26}
\end{gather}
Note that, we do not fix coupling function for the second Maxwell $f_B$, but derive it from the EOM with intent. Fixing, for example, $f_B =
f_1$ makes system \eqref{eq:2.05}--\eqref{eq:2.10} not self consistent,
therefore proper solution with the $f_B = f_1$ condition can't be
found. \\

\newpage
%\normalize

\end{document}